\documentclass[12pt]{iopart}

\usepackage{graphicx}
\usepackage{multirow}
\usepackage{dcolumn}
\usepackage{cite}

\begin{document}

\title[Unitary transformations of anharmonic oscillators]
{Unitary transformations of a family of two-dimensional anharmonic oscillators}

\author{Francisco M Fern\'andez \ and Javier Garcia}

\address{INIFTA (UNLP, CCT La Plata-CONICET), Divisi\'on Qu\'imica Te\'orica,
Blvd. 113 S/N,  Sucursal 4, Casilla de Correo 16, 1900 La Plata,
Argentina}

\ead{fernande@quimica.unlp.edu.ar}

\maketitle

\begin{abstract}
In this paper we analyze a recent application of perturbation theory by the
moment method to a family of two-dimensional anharmonic oscillators. By
means of straightforward unitary transformations we show that two of the
models studied by the authors are separable. Other is unbounded from below
and therefore cannot be successfully treated by perturbation theory unless a
complex harmonic frequency is introduced in the renormalization process. We
calculate the lowest resonance by means of complex-coordinate rotation and
compare its real part with the eigenvalue estimated by the authors. A pair
of the remaining oscillators are equivalent as they can be transformed into
one another by unitary transformations.
\end{abstract}

\section{Introduction}

\label{sec:intro}

Witwit and Killingbeck\cite{WK93} applied the perturbation theory by the
moment method developed by Fern\'{a}ndez and Castro\cite{FC84,AFMC84,FC85}
to a family two-dimensional anharmonic oscillators with ``mixed parity
potentials''. By means of renormalization of the perturbation series they
obtained reasonably accurate eigenvalues for a particular set of potential
parameters. The authors did not take into account the symmetry of the
perturbation which had proved to be extremely useful in the treatment of
similar quantum-mechanical models\cite{PE81a,PE81b}. Lately, the application
of point group symmetry (PGS) has proved to be suitable for the analysis of
a variety of non-Hermitian anharmonic oscillators\cite{FG14a,FG14b,AFG14a,
AFG14c}.

An attempt to apply PGS to the anharmonic oscillators studied by Witwit and
Killingbeck revealed some interesting facts that we want to discuss in this
paper. In section~\ref{sec:unitary_transf} we briefly summarize some aspects
of unitary transformations and point groups that will be useful in the
subsequent sections. In section~\ref{sec:anharmonic_osc} we apply those
concepts to the anharmonic oscillators studied by Witwit and Killingbeck.
Finally, in section \ref{sec:conclusions} we summarize the main results of
this paper and draw conclusions.

\section{Unitary transformations and point group}

\label{sec:unitary_transf}

If $U$ is an invertible operator, then the Hamiltonian operators $H$ and $%
\tilde{H}=UHU^{-1}$ are isospectral. In this paper we are interested only in
unitary transformations $U^{-1}=U^{\dagger }$, where $U^{\dagger }$ is the
adjoint of $U$.

The set of unitary transformations $U_{i}$, $i=1,2,\ldots ,h$ that leave a
given Hamiltonian operator $H$ invariant $U_{i}HU_{i}^{\dagger }=H$ form a
group with respect to the composition $U_{i}U_{j}$\cite{T64,C90}. It follows
from the invariance of $H$ that $[H,U_{i}]=0$. Clearly, if $\psi $ is an
eigenfunction of $H$ with eigenvalue $E$ then $U_{i}\psi $ is also
eigenfunction with the same eigenvalue as follows from $HU_{i}\psi
=U_{i}H\psi =EU_{i}\psi $. The eigenfunctions of $H$ are bases for the
irreducible representations (irreps) of the point group $G$ of $H$ and can
therefore be classified according to them\cite{T64,C90}.

If $G=\{U_{i},\;i=1,2,\ldots ,h\}$ is the point group for $H$ and $\tilde{H}%
=UHU^{\dagger }$ then $\tilde{G}=\{\tilde{U}_{i}=UU_{i}U^{\dagger
},\;i=1,2,\ldots ,h\}$ is the point group for $\tilde{H}$. Both groups are
isomorphic as follows from $\tilde{U}_{i}\tilde{U}_{j}=UU_{i}U_{j}U^{\dagger
}$. It the following section we apply these simple well known results to the
anharmonic oscillators studied by Witwit and Killingbeck.

\section{The anharmonic oscillators}

\label{sec:anharmonic_osc}

Witwit and Killingbeck\cite{WK93} applied perturbation theory by the moment
method (which they baptized inner product method) to anharmonic oscillators
of the form
\begin{eqnarray}
H &=&p_{x}^{2}+p_{y}^{2}+x^{2}+y^{2}+\lambda V(x,y)  \nonumber \\
V(x,y)
&=&a_{xx}x^{4}+4b_{xy}x^{3}y+6c_{xy}x^{2}y^{2}+4b_{yx}xy^{3}+a_{yy}y^{4},
\end{eqnarray}
where $\lambda $ is the perturbation parameter. They chose a few different
sets of potential parameters that we analyze in what follows:

Case 1: $a_{xx}=b_{xy}=c_{xy}=b_{yx}=a_{yy}=1$. It leads to
\begin{equation}
H=p_{x}^{2}+p_{y}^{2}+x^{2}+y^{2}+\lambda \left(
x^{4}+4x^{3}y+6x^{2}y^{2}+4xy^{3}+y^{4}\right) .  \label{eq:H_C1a}
\end{equation}
The unitary transformation
\begin{equation}
U:(x,y)\rightarrow \left( \frac{x}{\sqrt{2}}-\frac{y}{\sqrt{2}},-\frac{x}{%
\sqrt{2}}-\frac{y}{\sqrt{2}}\right) ,  \label{eq:Ch_var_Case1}
\end{equation}
decouples the degrees of freedom and leads to
\begin{equation}
\tilde{H}=p_{x}^{2}+p_{y}^{2}+x^{2}+y^{2}+4\lambda y^{4},  \label{H_C1b}
\end{equation}
facilitating the calculation enormously. For instance, a straightforward
application of the Riccati-Pad\'{e} method (RPM)\cite{FMDT89, FMT89b} for $%
\lambda =1$ yields
\begin{equation}
E_{0}^{C1}=2.903136945459000022293850722201023931817
\end{equation}
that considerably improves the estimate of Witwit and Killingbeck\cite{WK93}.

Case 2: $a_{xx}=c_{xy}=a_{yy}=1$, $b_{xy}=b_{yx}=0$. This problem
\begin{equation}
H=p_{x}^{2}+p_{y}^{2}+x^{2}+y^{2}+\lambda \left(
x^{4}+6x^{2}y^{2}+y^{4}\right)   \label{eq:H_C2a}
\end{equation}
is also separable by means of the unitary transformation
\begin{equation}
U:(x,y)\rightarrow \left( \frac{x}{\sqrt{2}}+\frac{y}{\sqrt{2}},\frac{y}{%
\sqrt{2}}-\frac{x}{\sqrt{2}}\right) ,  \label{eq:Ch_var_Case2}
\end{equation}
that leads to
\begin{equation}
\tilde{H}=p_{x}^{2}+p_{y}^{2}+x^{2}+y^{2}+2\lambda \left( x^{4}+y^{4}\right)
.  \label{eq:H_C2b}
\end{equation}
The RPM for the ground state of this problems with $\lambda =10^{6}$ yields
\begin{equation}
E_{0}^{C2}=267.2002503791361424618416691534920465128,
\end{equation}
which also improves the result reported by those authors.

Case 3: $a_{xx}=a_{yy}=0$, $c_{xy}=b_{xy}=b_{yx}=1$. The resulting
anharmonic oscillator
\begin{equation}
H=p_{x}^{2}+p_{y}^{2}+x^{2}+y^{2}+\lambda \left(
4x^{3}y+6x^{2}y^{2}+4xy^{3}\right)  \label{eq:H_C3a}
\end{equation}
is obviously unbounded from below and does not support bound states. For
this reason the results of Witwit and Killingbeck\cite{WK93} for this
example were considerably less accurate than for the other ones. They did
not mention this fact and did not report results for $\lambda >0.12$,
probably because the increasingly greater imaginary part of the resonances
made the straightforward perturbation calculation unreliable. In principle,
one can obtain a convergent perturbation series if the harmonic frequency
introduced for renormalization is allowed to be complex instead of being
restricted to real values.

This problem exhibits symmetry $C_{2v}$ and the same change of variables
shown above in equation (\ref{eq:Ch_var_Case2}) leads to
\begin{equation}
\tilde{H}=p_{x}^{2}+p_{y}^{2}+x^{2}+y^{2}+\lambda \left( \frac{7y^{4}}{2}%
-3x^{2}y^{2}-\frac{x^{4}}{2}\right) ,  \label{eq:H_C3b}
\end{equation}
with obviously the same symmetry $C_{2v}$. However, since this potential
exhibits only even powers of the variables the authors could have simplified
the calculation by resorting to the simple symmetry analysis already used in
earlier papers\cite{KJ86,W91b}. For a more rigorous discussion of the
interplay between symmetry and perturbation theory by the moment method for
this kind of anharmonic oscillators see a recent paper by Fern\'{a}ndez\cite
{F14}.

We calculated the lowest resonance for the anharmonic oscillator (\ref
{eq:H_C3a}) by means of the complex rotation method\cite
{BC71,R76,CR77,YBLBF78} using finite basis sets of eigenfunctions of $H_{0}$
of increasing dimension up to $30^{2}\times 30^{2}$ and a roughly optimal
rotation angle $\theta =0.06\pi $. Table~\ref{tab:E_C3} shows present
results and those of Witwit and Killingbeck\cite{WK93} for some values of $%
\lambda $. Note that the error in the eigenvalue estimated by those authors
is of the order of $|\Im E|$ that increases with $\lambda $ as argued above.
We have chosen the greatest values of $\lambda $ considered by those authors
in order to illustrate this point more clearly.

Case 4: $a_{xx}=a_{yy}=c_{xy}=1$, $b_{xy}=b_{yx}=-1$. The inclusion of this
case in the discussion is surprising because the resulting Hamiltonian
becomes the one for Case 1 by means of the unitary transformations $%
U:(x,y)\rightarrow (-x,y)$ or $U:(x,y)\rightarrow (x,-y)$. Note that the
authors obtained exactly the same eigenvalues with the same accuracy for
both cases as expected. There is no point in discussing this case here.

Case 5: $c_{xy}=1$, $a_{xx}=a_{yy}=b_{xy}=b_{yx}=0$. The resulting
Hamiltonian
\begin{equation}
H=p_{x}^{2}+p_{y}^{2}+x^{2}+y^{2}+6\lambda x^{2}y^{2},
\end{equation}
does not exhibit ``mixed parity'' but symmetry $C_{4v}$ \cite{PE81a} and has
been chosen as benchmark many times in the past (other references are given
elsewhere\cite{AF10}) even for an earlier application of perturbation theory
by the moment method\cite{KJ86,W91b}.

\section{Conclusions}

\label{sec:conclusions}

The aim of the addendum by Witwit and Killingbeck\cite{WK93} was the
application of the perturbation theory by the moment method to
two-dimensional oscillators with mixed-parity potentials; that is to say:
with even and odd powers of the variables $x$ and $y$. However, present
analysis shows that two of the models studied by those authors (Case 1 and
Case 2) are trivial in the sense that they are separable by unitary
transformations. What is more, the resulting Hamiltonians exhibit only even
powers of the coordinates. The Hamiltonian operator for Case 3 is unbounded
from below and therefore cannot be successfully treated by perturbation
theory unless a complex harmonic frequency is introduced in the
renormalization process. Besides, this Hamiltonian can be transformed into
one with only even powers of the coordinates. The Hamiltonian for Case 4 is
trivially isospectral to the one for Case 1 and, therefore, does not add
anything relevant to the discussion. Finally, the Hamiltonian for Case 5
does not exhibit mixed parity and was treated before by the same authors\cite
{KJ86,W91b} (see Fern\'{a}ndez\cite{F14} for a discussion based on PGS).

\ack This report has been financially supported by PIP No.
11420110100062 (Consejo Nacional de Investigaciones Cientificas y
Tecnicas, Rep\'{u}blica Argentina)

\begin{table}[H]
\caption{Lowest resonance for the anharmonic oscillator ( \ref{eq:H_C3a})}
\label{tab:E_C3}
\begin{center}
\par
\begin{tabular}{llll}
\hline
$\lambda$ & Ref.\cite{WK93} & \multicolumn{1}{c}{$\Re E$} &
\multicolumn{1}{c}{$\Im E$} \\ \hline
0.10 & 2.0733 & 2.07335064 & -0.000459014 \\
0.12 & 2.08 & 2.0746525 & -0.0022857 \\
0.13 &  & 2.0738983 & -0.0041665 \\
0.14 &  & 2.0724187 & -0.0068909 \\ \hline
\end{tabular}
\end{center}
\end{table}


\begin{thebibliography}{99}
\bibitem{WK93}  Witwit N R M and Killingbeck J 1993 \textit{J. Phys. A}
\textbf{26} 3659.

\bibitem{FC84}  Fern\'{a}ndez F M and Castro E A 1984 \textit{Int. J.
Quantum Chem.} \textbf{26} 497.

\bibitem{AFMC84}  Arteca G A, Fern\'{a}ndez F M, Mes\'{o}n A M, and Castro E
A 1984 \textit{Physica A} \textbf{128} 253.

\bibitem{FC85}  Fern\'{a}ndez F M and Castro E A 1985 \textit{Int. J.
Quantum Chem.} \textbf{28} 603.

\bibitem{PE81a}  Pullen R A and Edmonds A R 1981 \textit{J. Phys. A} \textbf{%
14} L477.

\bibitem{PE81b}  Pullen R A and Edmonds A R 1981 \textit{J. Phys. A} \textbf{%
14} L319.

\bibitem{FG14a}  Fern\'{a}ndez F M and Garcia J 2014 \textit{Ann. Phys.}
\textbf{342} 195.

\bibitem{FG14b}  Fern\'{a}ndez F M and Garcia J 2014 \textit{J. Math. Phys.}
\textbf{55} 042107.

\bibitem{AFG14a}  Amore P, Fern\'{a}ndez F M, and Garcia J 2014 \textit{Ann.
Phys.}, in the press.

\bibitem{AFG14c}  Amore P, Fern\'{a}ndez F M, and Garcia J 2014
Non-Hermitian oscillators with Td symmetry arXiv:1409.2672 [quant-ph]

\bibitem{T64}  Tinkham M 1964 \textit{Group Theory and Quantum Mechanics}
(McGraw-Hill Book Company, New York).

\bibitem{C90}  Cotton F A 1990 \textit{Chemical Applications of Group Theory}
(John Wiley \& Sons, New York).

\bibitem{FMT89b}  Fern\'{a}ndez F M, Ma Q, and Tipping R H 1989 \textit{%
Phys. Rev. A} \textbf{40} 6149.

\bibitem{FMDT89}  Fern\'{a}ndez F M, Ma Q, DeSmet D J, and Tipping R H 1989
\textit{Can. J. Phys.} \textbf{67} 931.

\bibitem{KJ86}  Killingbeck J and Jones M N 1986 \textit{J. Phys. A} \textbf{%
19} 705.

\bibitem{W91b}  Witwit N R M 1991 \textit{J. Phys. A} \textbf{24} 4535.

\bibitem{F14}  Fern\'{a}ndez F M 2014 Perturbation theory by the moment
method and point-group symmetry. arXiv:1409.4120 [quant-ph].

\bibitem{BC71}  Balslev E and Combes J C 1971 \textit{Commun. Math. Phys.}
\textbf{22} 280.

\bibitem{R76}  Reinhardt W P 1976 \textit{Int. J. Quantum Chem. Symposium}
\textbf{10} 359.

\bibitem{CR77}  Chu S-I and Reinhardt W P 1977 \textit{Phys. Rev. Lett.}
\textbf{39} 1195.

\bibitem{YBLBF78}  Yaris R, Bendler J, Lovett R A, Bender C A, and Fedders P
A 1978 \textit{Phys. Rev. A} \textbf{18} 1816.

\bibitem{AF10}  Amore P and Fern\'{a}ndez F M 2010 \textit{Phys. Scr.}
\textbf{81} 045011.
\end{thebibliography}
\end{document}